# Imaging, counting, and positioning single interstitial atoms in solids

**Authors:** Jizhe Cui[1,2,3†], Haozhi Sha[1,2,3†], Liangze Mao[1,2,3], Kang Sun[1,2,3], Wenfeng Yang[1,2,3] and Rong Yu[1,2,3*]

Affiliations:

[1]School of Materials Science and Engineering, Tsinghua University, Beijing 100084, China.

[2]Key Laboratory of Advanced Materials of Ministry of Education, Tsinghua University, Beijing 100084, China.

[3]State Key Laboratory of New Ceramics and Fine Processing, Tsinghua University, Beijing 100084, China.

*Corresponding author. Email: ryu@tsinghua.edu.cn

†These authors contributed equally to this work.

**Abstract**

**Interstitial atoms are ubiquitous in solids and they are widely incorporated into materials to tune their lattice structure, electronic transportation, and mechanical properties. Because the distribution of interstitial atoms in matrix materials is usually disordered and most of them are light atoms with weak scattering ability, it remains a challenge to directly image single interstitial atoms and measure their geometrical positions. In this work, direct imaging and measuring of single interstitial atoms have been realized with adaptive-propagator ptychography. The measurement of their three-dimensional coordinates enables quantitative analysis of the pair distribution function of the interstitial atoms and reveals the anisotropic occupation of oxygen in the interstitial sites in titanium. The current work paves the way for the determination of interstitial atoms in materials, and for the correlation between the atomic-scale behavior of interstitial atoms and the physical properties of materials.**

**Introduction**

Interstitial atoms are ubiquitous in materials and play important role in tuning the mechanical and physical properties of semiconductors, superconductors, and alloys [1-4]. Recently, incorporation of interstitial atoms in alloys has been increasingly employed to improve mechanical properties of alloys [5-11]. As a paradigm of materials science, understanding the physical properties often requires studying the atomic structure of the materials [12,13]. Therefore, it is imperative to obtain atomic-scale information of interstitial atoms for understanding and controlling the behavior of interstitial atoms in materials. Previous studies on interstitial atoms mostly rely on statistical methods, such as the positron annihilation spectroscopy [14], X-ray diffuse scattering [15], internal friction [16] and electron diffraction[17]. While these methods are useful, they often sacrifice spatial resolution and provide only average results.

Direct imaging of interstitial atoms in solids has been challenging due to two primary reasons. First, interstitial atoms are usually light atoms, like carbon, nitrogen, and oxygen; they give very weak electron scattering compared with the matrix atoms. Second, the surfaces of solids are often structurally modified, damaged, or contaminated, forming a layer with different atomic structure than the bulk inside. The electron scattering signals from the surface layers result in strong noise that prevent the detection of interstitial atoms inside the bulk. As a result, direct imaging of interstitial atoms in solids requires very high signal-to-noise ratio in images and the elimination of the surface contributions to the bulk signals.

Recently, oxygen atoms in a TiZrNb alloy of the body-center-cubic structure was shown to occupy the tetrahedral interstitial sites [18]. Since the concentrations (12 at.%) of oxygen in the alloy is high, the interstitial atoms are crowded together with an average spacing of 6 Å. These densely distributed interstitial atoms can provide stronger signals for detection, but at the same time make the detection only provide their average information and not about the behavior of individual interstitial atoms.

Interstitial atoms usually have low concentrations in solids; even a few parts per million of interstitial doping is enough to significantly improve properties [1,19]. It is therefore important to be able to detect single interstitial atoms, even though they can only give weak signals for detection. Here we show that imaging, counting, and positioning of single interstitial atoms in lightly alloyed solids is possible using adaptive propagator ptychography (APP) [20]. Slightly oxidized pure titanium is used as a model system. We revealed that interstitial O atoms display an anisotropic occupation in the interstitial sites in the matrix, rather than a random arrangement, even when the atomic concentration is below 1%.

**Results and discussions**

In this study, we investigated the behavior of interstitial oxygen in α-Ti. Using the sample preparation methods described in the Methods section, we heat-treated Pure-phase α-Ti in an oxygen atmosphere. The diffusion of oxygen into the bulk results in a gradient of oxygen in the matrix with interstitial atoms of continuously varying concentration, providing a convenient platform for the current study, as illustrated in **Fig. 1**a. The direction of the arrow indicates the direction of reducing oxygen content. The labeled regions 1-4 correspond to the areas where ptychographic reconstruction and analysis were performed.

Electron ptychography is a computational imaging method that combines scanning transmission electron microscopy and coherent diffraction imaging (CDI) [21,22]. It is characterized by high dose efficiency, high spatial resolution, high phase accuracy, and depth resolution [23-25]. By using multislice electron ptychography with an adaptive propagator [20], deep-sub-angstrom resolution for defective materials [24], sub-angstrom resolution for beam-sensitive materials [26], and lattice-resolved antiferromagnetic imaging [25] have been realized. The multislice electron ptychography with adaptive propagator enables direct imaging of interstitial atoms, thanks to several crucial features. First, it eliminates aberrations during the imaging process, thus improving image quality over aberration-corrected electron microscopy. Second, it is capable of obtaining high accuracy in the retrieved phase of the electron wave function, making it possible to detect weak signals from interstitial atoms. Third, it has good depth-resolution in the electron beam direction and helps to remove signal interference from surface damage or contamination layers, which is particularly crucial for the observation of discrete and weak interstitial signals. Taking an interstitial oxygen atom in a 20-nm-thick sample as an example. In the projected phase of the 20 nm sample, the interstitial O atom accounts for a proportion of 0.4%, whereas in the 1 nm thick slice, it accounts for a proportion of 8%, much higher than 0.4%. Therefore, the employment of 'multislice' significantly enhances the detectability of single interstitial atoms.

**Fig. 1** presents a schematic to the experiment analysis, including the sample, the APP method, and the reconstructed phase images. Using Area 4 as an illustrative example, we utilize the STEM mode to perform a point-by-point scan using a convergent electron beam. A direct electron detector (EMPAD) [27] captures the diffraction pattern at each scanning point, as shown in **Fig. 1**b. The datasets collected serve as input for the multislice APP method, enabling iterative reconstruction and generating a sequence of object slices, as depicted in **Fig. 1**c. After that, we can effectively eliminate damages and oxidations present on the upper and lower surfaces of the sample (shown in **Fig. 1**e and **1**f). Consequently, we are able to analyze intermediate slices that describe the intrinsic structure inside the bulk.

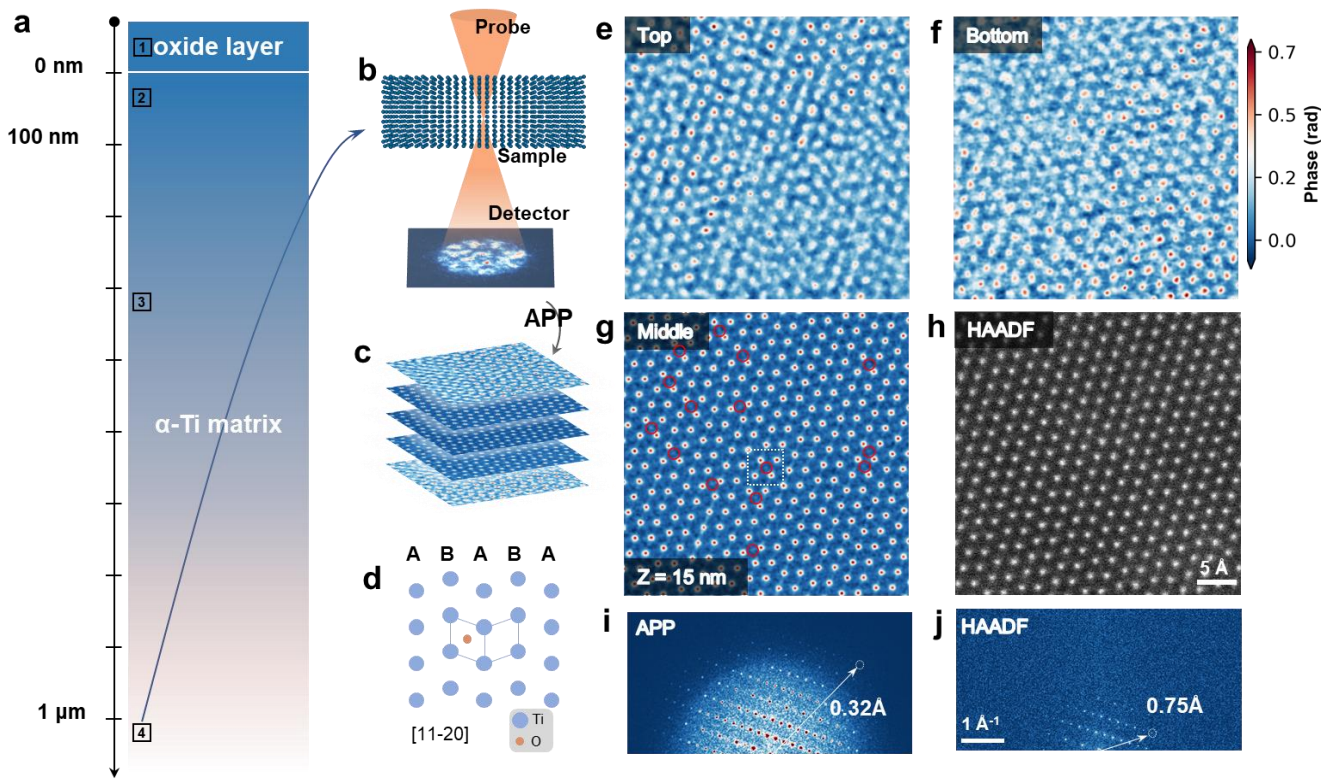

**Fig. 1. The schematic of the procedure of investigating single interstitial atoms.** (a)

Schematic representation of a cross-section FIB sample obtained from oxidized pure titanium metal. The arrow indicates the direction of oxygen content reduction. The labeled regions 1-4 correspond to the four regions where the APP reconstruction was performed. (b) Conceptual diagram illustrating the process of 4D data acquisition. (c) Slices showing the reconstructed phase images obtained through the multislice APP. Each slice represents sample information at different depths along the direction of the electron beam. (d) Atomic arrangement of the [11-20] zone axis of α-Ti. The orange atoms represent interstitial oxygen atoms occupying octahedral interstices. The stacked atomic layers are denoted as A and B. (e, f, and g) Three typical slices of reconstructed phase images of region 4, where surface damages could be found in the top and bottom layers. (h) The corresponding HAADF image. The phase image (g) clearly exhibits interstitial oxygen atoms located at the octahedral interstices, highlighted with red circles. In (h), the HAADF image only displays the matrix Ti atoms, without observable interstitial atoms. (i) The diffractogram corresponding to (g), showing an information limit of 0.32 Å. (j) The diffractogram of the HAADF image, showing an information limit of 0.75 Å.

Gradient distribution of oxygen diffusion in the matrix

APP reconstruction was performed for a series of regions depicted in **Fig. 1**a, with varying oxygen contents. Combining the EDS results (shown in Extended Data Fig. 1) and the HAADF images (displayed in Extended Data Fig. 2), it is evident that region 1 corresponds to the oxide layer and regions 2-4 represent matrix regions that maintain the α-Ti structure. The EDS signals of Extended Data Fig. 1 indicate lower oxygen content in these regions, which decreases gradually with depth. It seems that even deep inside the bulk, the concentration of O remains higher than 10%, which comes from the surface oxidation of the thin TEM sample. It illustrates again the necessity of utilizing multislice ptychography for the study of intrinsic behavior of interstitial atoms inside the bulk.

Extended Data Fig. 2 shows the average phase of regions 2-4. There are numerous interstitial atoms in region 2 at the octahedral sites indicated in **Fig. 1**d, suggesting high oxygen content at this depth. In region 3, the content of interstitial atoms is noticeably lower than that in region 2. In region 4, the contrast of interstitial atoms becomes weak and difficult to detect. For convenience, we designate region 2 as the High-O area, region 3 as the Mid-O area, and region 4 as the Low-O area.

Direct imaging of single interstitial oxygen

While the damaged surface layers (**Fig. 1**e and **1**f) are eliminated, weak image spots at octahedral sites can be readily observed in the intermediate slices, as highlighted by red circless in **Fig. 1**g. As a comparison, only Ti atoms are observable in the HAADF image (**Fig. 1**h) of the same regions. Through quantitative analysis, we will show that the weak spots in the phase images are oxygen interstitial atoms. **Fig. 1**i and 1j show the diffractograms of an APP phase image and a HAADF image, respectively. The information limit of the APP phase is much higher than that of HAADF, supporting direct observation of single interstitial atoms using the APP method.

**Fig. 2** compares the experimental and simulated phase images of single oxygen interstitial atoms in α-Ti. For the simulation dataset, the thickness of Ti is 20 nm, with an O atom occupying an octahedral interstitial site. After multislice electron diffraction simulation and APP reconstruction, the phase images are shown in **Fig. 2**a-c. As shown in **Fig. 2**a, the signal of the interstitial atom was only visible at the depth of 7 nm, but not at 2 nm and 12 nm. Due to the limited depth resolution in the electron beam direction, the interstitial atom is elongated into a spindle in the depth profile of the reconstructed phase images, as shown in **Fig. 2**b. **Fig. 2**c show the lateral profile cross the center of the spindle in the depth profile. The interstitial oxygen atom is marked. It is worth noting that spots in individual phase images do not necessarily correspond to interstitial atoms; they are possibly noise spots. Only spindles in depth profiles correspond to interstitial atoms.

The experimental phase images at various depths are shown in **Fig. 2**d, which corresponds to the outlined box in **Fig. 1**g. The depth profile along the dashed line in **Fig. 2**d is shown in **Fig. 2**e, revealing a spindle for an interstitial oxygen. A lateral profile across the center of the spindle is shown in **Fig. 2**f, demonstrating the presence of a peak corresponding to the interstitial oxygen.

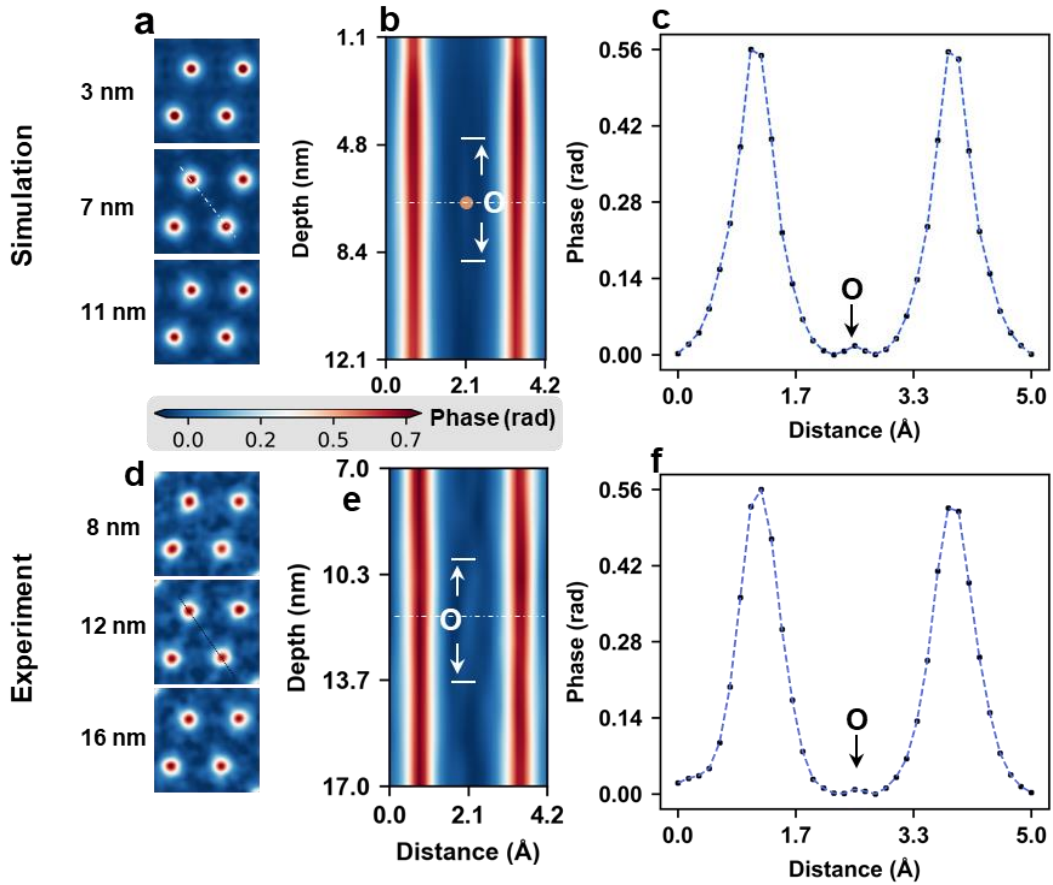

**Fig. 2. Direct imaging of a single interstitial oxygen.** (a, b, and c) A single interstitial oxygen in APP phase images reconstructed using the simulated dataset. (a) the slices at different depth, the single interstitial oxygen can only be seen at the depth of 7 nm. (b) The depth profile along the dashed line in (a), showing the spindle shape of the interstitial atom. The true position of the interstitial oxygen atom is at the center of the spindle, as indicated by the orange spot. (c) The lateral profile across the center of the spindle in (b), with the peak of the interstitial oxygen indicated. (d, e, and f) The experimental results of an interstitial oxygen atom. (d) phase images at various depths. An interstitial O atom is at 11.6 nm. (e) Depth profile along the dashed line in (d). (f) The lateral profile across the center of the spindle in (e), with the interstitial O indicated.

Positioning interstitial atoms

As shown above, interstitial atoms appear spindles in depth profiles of the phase images (**Fig. 2**b and **2**e) due to the limited resolution along the electron beam direction (about 2-5 nm). In **Fig. 2**b, the true position of the interstitial atom is indicated, and the center of the spindle matches well the true position. Therefore, the position of the interstitial atom in the electron beam direction can be determined by measuring the center of the spindle.

In the [11-20] direction of α-Ti, the atomic spacing is 2.95 Å, which is far smaller than the depth resolution. Consequently, the elongation of atoms due to the limited depth resolution hinders the determination of three-dimensional (3D) configuration of matrix Ti atoms. Fortunately, the determination of 3D configuration of interstitial atoms is possible at the limited depth resolution. This is because the content of interstitial atoms is typically below 1%. Roughly estimating, for randomly distributed interstitial atoms, the spacing between two interstitial atoms along the same atomic column is about 10 nm, much greater than the depth resolution (2.9 nm in this study). The position of an interstitial atom in the depth direction corresponds to the center of the spindle, as indicated by the dashed line in **Fig. 2**e (at 11.6 nm).

We first performed simulation test. Fifty oxygen atoms are randomly located at octahedral sites in a supercell of 6 nm × 6 nm × 20 nm (with 20 nm along the [11-20] direction), corresponding to a content of 0.18 at.%. A 4D dataset was generated through multislice simulations and the phase images were reconstructed using the APP method. Extended Data Fig. 3 错误!未找到引用源。 displays the 3D distribution of the interstitial oxygen determined according to above procedure, together with the true positions in the model. All 50 interstitial oxygen atoms are successfully identified.

3D distribution of interstitial oxygen

In the low-O region (region 4) in **Fig. 1**a, 144 interstitial atoms are identified, corresponding to a content of 0.3 at.%. The interstitial atoms are displayed in **Fig. 3**a, with their projections shown in **Fig. 3**b-d. The distribution of interstitial oxygen in the Low-O region seem random and is relatively uniform in the image plane. The uneven distribution in the depth direction may be related to the formation of the surface oxide layers.

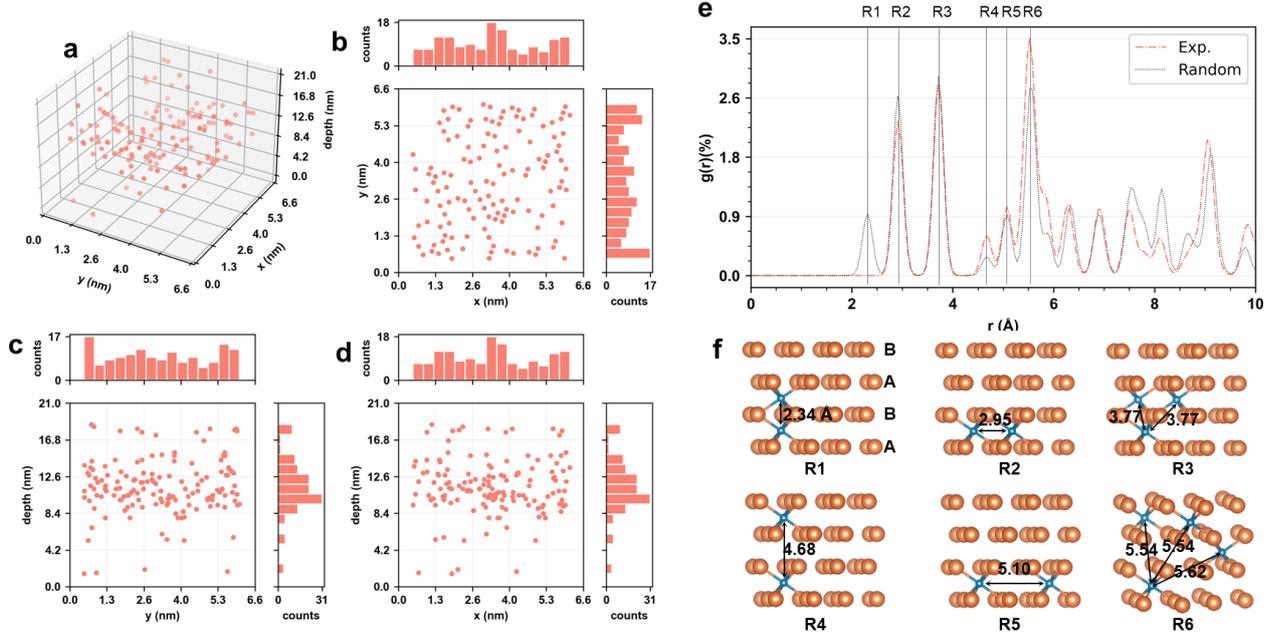

**Fig. 3. 3D distribution of interstitial oxygen in the Low-O region.** (a) The 3D distribution of interstitial oxygen atoms obtained in this region. (b, c, and d) Projections along the three axes in (a). Each projection distribution graph contains a histogram of the number of interstitial atoms projected onto two coordinate axes. (e) The pair distribution function g(r) of the interstitial atoms in (a). The x-axis represents the radial distance r between atoms, and the y-axis represents the corresponding proportion of atom pairs at that distance. The labels R1-R6 indicate the distances between interstices, as shown in (f): R1 = 2.34 Å, R2 = 2.95 Å, R3 = 3.77 Å, R4 = 4.68 Å, R5 = 5.10 Å. R6 represents a collection of distances of 5.54 Å and 5.62 Å. In (e), the black dotted line represents the PDF curve of completely disordered octahedral interstices, while the red dashed line represents the experimental result of interstitial oxygen.

The pair distribution function (PDF) g(r) (where r represents the distance between interstitials) of the measured oxygen atoms was calculated and compared with that for hypothetical randomly-distributed interstitial atoms, as shown in **Fig. 3**e. The atomic distances of the PDF peaks and the corresponding atomic environments are depicted in **Fig. 3**f.

We note that the experimental PDF curve and that for the hypothetical randomly-distributed interstitial atoms do not coincide, indicating that the distribution of interstitial O in α-Ti is not completely disordered. The most obvious difference is the absence of the R1 peak in the experimental PDF. The R1 peak corresponds to the atomic configuration with neighboring oxygen atoms in the *c* direction occupying the same in-plane position (**Fig. 3**f-R1). The absence of the R1 peak indicates that the occupancy of neighboring octahedral sites in the $c$ direction would lead to high energy and is not allowed. The presence of the R2 peak in the experimental results implies that the neighboring octahedral sites in the basal plane are allowed to be occupied simultaneously. Finally, the absence of the R1 peak and the presence of the R2 and R4 peaks indicate a medium-range ordering (MRO) of interstitial O in the α-Ti. In other words, in a local region, interstitial O atoms tends to occupy interstitial sites within the same basal plane, forming a planar configuration and leaving the interstitial sites in neighboring planes vacant. This phenomenon will be further demonstrated later in local strain analysis.

Interstitial-induced anisotropic lattice expansion

In the Mid-O region (Region 3 in **Fig. 1**a), 1000 interstitial atoms are identified, corresponding to a content of 1.1 at.%. As described above, the interstitial O atoms are not completely disordered even in the Low-O region; they avoid occupying neighboring sites in the $c$ direction. As the oxygen content increases, the interstitial atoms form higher degree of ordering.

**Fig. 4a** shows two phase images of the Mid-O region at the depth of 7 nm and 17 nm. Alternating basal planes are labelled by blue and yellow rhombuses. As indicated by the bule and yellow frames, the interstitial atoms form planar configuration in the basal planes, which are occupied alternatively. The size of the interstitial platelets is nanometer scale. It is interesting that the occupation would change with depth. At the depth of 7 nm, interstitial atoms occupy the basal plane labelled by the blue rhombus. At the depth of 17 nm, they occupy the basal plane labelled by the yellow rhombus.

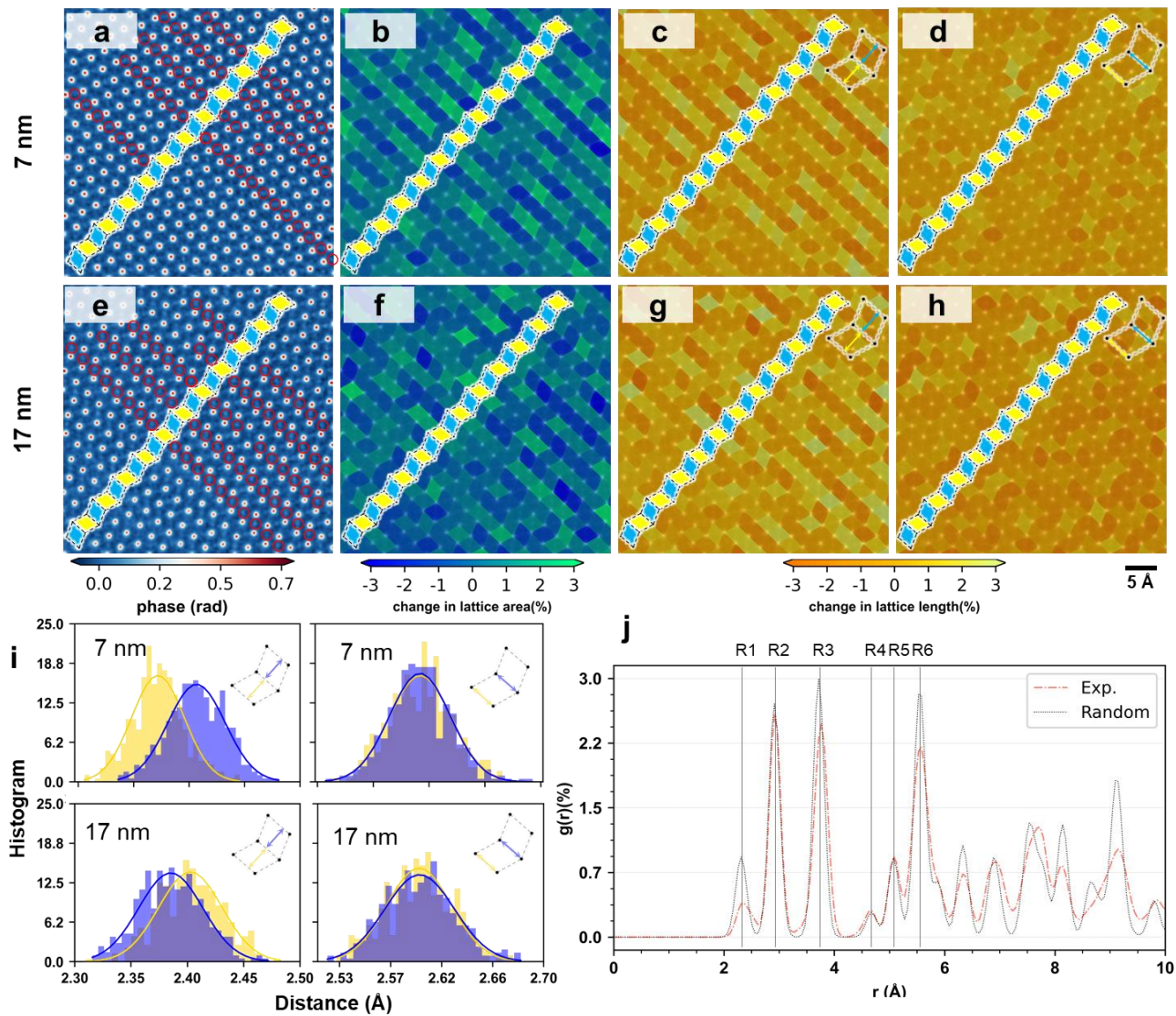

**Fig. 4. The medium-range ordering of interstitial oxygen in Mid-O region.** (a) Phase image at a depth of 7 nm, with alternating basal planes labeled by blue and yellow rhombuses. Interstitial atoms are located in the basal planes marked by the blue rhombuses, and typical interstitial sites are highlighted with red circles. (b, c, and d) Represent changes in lattice area, variations in the distance between adjacent basal planes, and changes in lattice length within the basal plane, respectively, illustrating anisotropic lattice expansion induced by the interstitials. (e, f, g, and h) Show the phase image and lattice expansion at a depth of 17 nm, where interstitial oxygen atoms occupy the basal planes labeled by the yellow rhombuses. (i) Histograms of the lengths shown in (c, d, g, and h), indicating a bimodal distribution of distances between adjacent basal planes but a unimodal distribution of lattice length within the basal plane at both 7 nm and 17 nm depths. The introduction of interstitial atoms primarily induces deformation perpendicular to the basal plane. (j) The PDF curves of interstitial oxygen in the Mid-O region, with the atomic distribution in three dimensions used for calculating the PDF presented in Extended Data Fig. 4.

The formation of interstitial platelets is associated with anisotropic lattice expansion, which is about 2% in the *c* direction (**Fig. 4**c and **5**g), but almost zero inside the basal planes (**Fig. 4**d and **5**h). **Fig. 4**i shows the histograms of the interplanar spacings as indicated by the insets.

The PDF curve of the interstitial oxygen atoms in the Mid-O region is shown in **Fig. 4**j. A minor peak is observed at the R1 position within this region. This appearance of the R1 peak indicates that a small number of O atoms are forced into unfavorable positions due to the increased O content. **Fig. 4**j reveals a noticeable rightward shift of the R1 and R3 peaks, while the R2 peak remains unaffected, which is consistent with the anisotropic lattice expansion.

**Conclusions**

In summary, we have realized direct imaging, counting, and positioning of single interstitial O atoms in titanium through adaptive propagation ptychography. Based on the spindle feature of the interstitial atoms in the depth direction, the coordinate of the interstitial atoms can be measured. Quantitative analysis indicates that the interstitial O atoms in α-Ti form local ordering, which depends on the oxygen content. For 0.3% content of oxygen, the interstitial atoms avoid occupying neighboring sites in the *c* direction. For 1.1% content of oxygen, they form nanoscale platelets with anisotropic lattice expansion. Imaging, counting, and positioning of single interstitial atoms pave the way for a deeper understanding of the behavior of interstitial atoms and fine tuning of the physical, mechanical, and chemical properties of materials.

**Materials and methods**

Materials and Sample preparation

Pure-phase α-Ti was placed in a tube furnace for heating. Throughout the process, oxygen was introduced at a pressure of 0.2 atmospheres, and the sample was maintained at a temperature of 500 °C for one hour to facilitate the diffusion of interstitial oxygen. A thin oxide layer, approximately 50 nm thick, formed on the surface of the Ti metal. Oxygen diffusion into the bulk produces a gradient of oxygen concentration within the matrix.

Experimental data collection and reconstruction

The experimental 4D datasets were collected using a probe-aberration-corrected FEI Titan Cubed Themis G2 electron microscope, equipped with a pixelated detector known as EMPAD. The accelerated voltage applied was 300 kV, and the convergence semi-angle was set to 25 mrad. Each diffraction pattern in the dataset had dimensions of 128×128, with a pixel size of 0.342 1/Å. The probe was under-focused by approximately 20 nm, with a scan step size of 0.468 Å. The dataset was collected using a beam current of roughly 20 pA and a dwell time of 0.5 ms. The diffraction patterns were padded to a size of 200×200 during the reconstruction, corresponding to a pixel size of 0.146 Å in real space.

Simulation of ptychography dataset

We used a custom-built multislice simulation program to generate the 4D dataset for ptychographic reconstruction. The program's reliability has been confirmed in our previous studies. We ensured that the simulation parameters—voltage, convergence angle, and pixel size of the diffraction patterns—matched those used in the actual experiments. The simulated model had a total thickness of 20 nm, closely resembling the experimental sample. To mimic experimental conditions, Poisson noise was added to the diffraction patterns, corresponding to a beam current of 20 pA and a dwell time of 0.5 ms.

**Acknowledgments:**

In this work we used the resources of the Physical Sciences Center and Center of High-Performance Computing, Tsinghua University.

**Funding:**

This work was supported by the National Natural Science Foundation of China (52388201 and 51525102).

**Author contributions:**

R.Y. designed and supervised the research. J.C. and H.S. performed simulations, experiments, and reconstructions. R.Y. and J.C. co-wrote the manuscript. All authors performed data analysis, discussed the results.

**Competing interests:**

The authors declare no competing financial interests.

**Extended Data Figures**

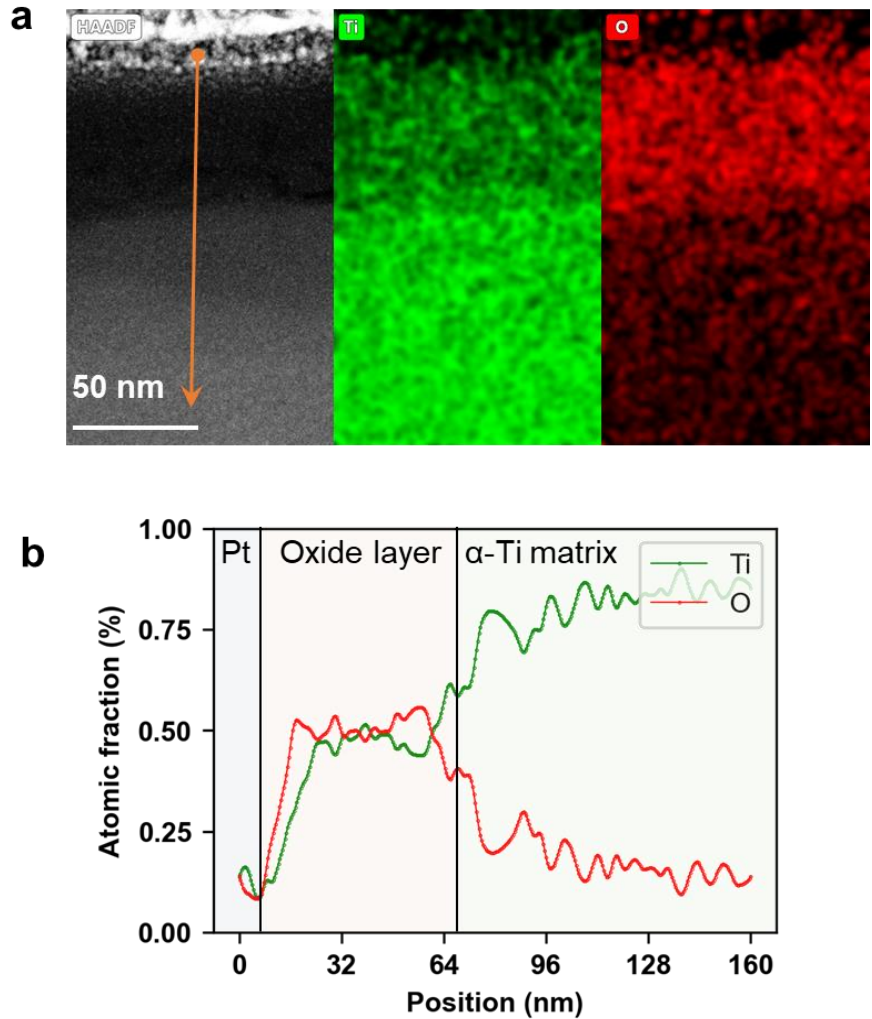

**Extended Data Fig. 1. EDS result of FIB sample.** The EDS results from the sample's oxide layer to the interior of the bulk sample are shown. (a) the EDS mapping distribution results, revealing the clear interface between the oxide layer and the α-Ti matrix. (b) the line scan EDS results along the orange line marked in (a). In the region of the oxide layer, the atomic ratio of Ti: O is close to 1: 1, and as we go deeper into the bulk sample, the oxygen content stabilizes at around 18%. This contribution of oxygen percentage is primarily due to the surface oxidation of the transmission electron microscopy (TEM) sample cut by FIB. In the experiment, it is necessary to remove the oxide layer on the upper and lower surfaces of the FIB sample to obtain the real information of the sample.

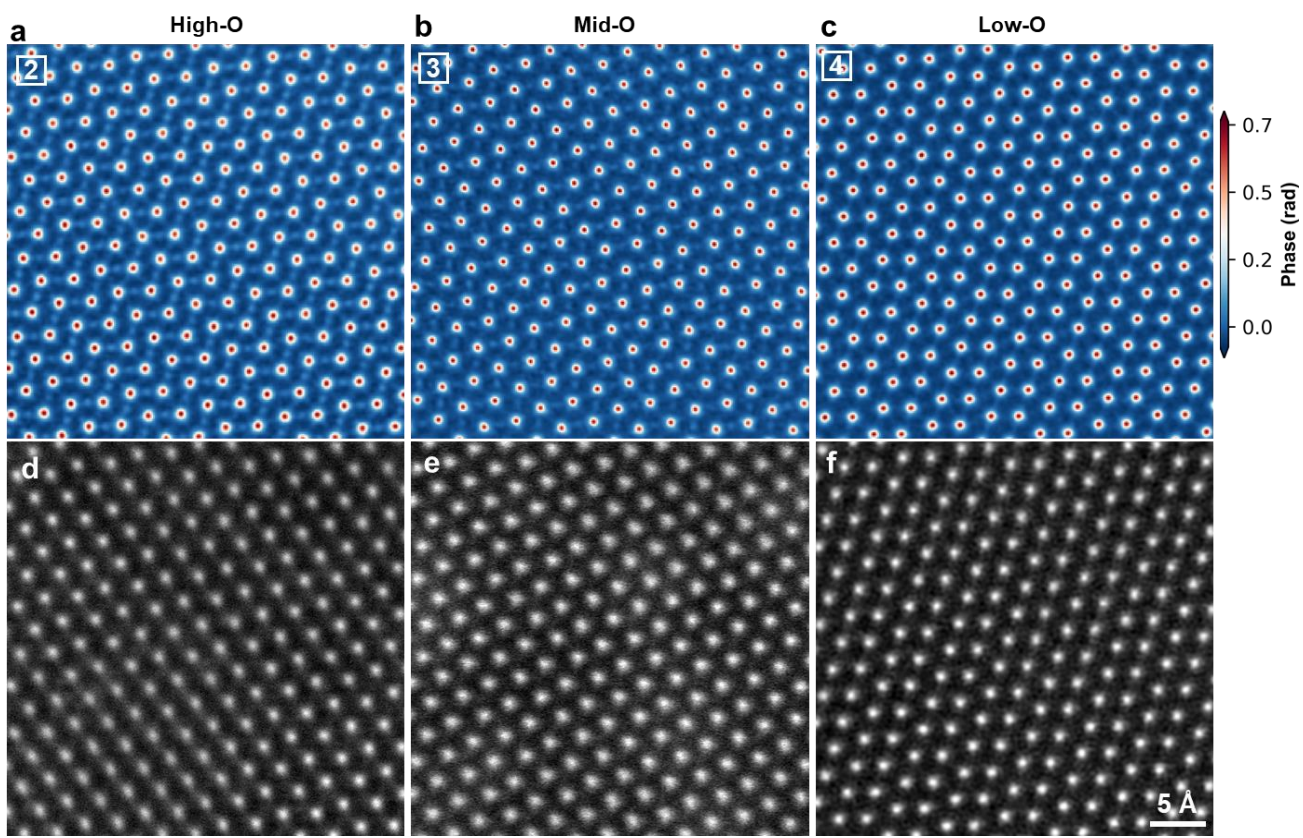

**Extended Data Fig. 2. APP results of different regions revealing the different O content.**

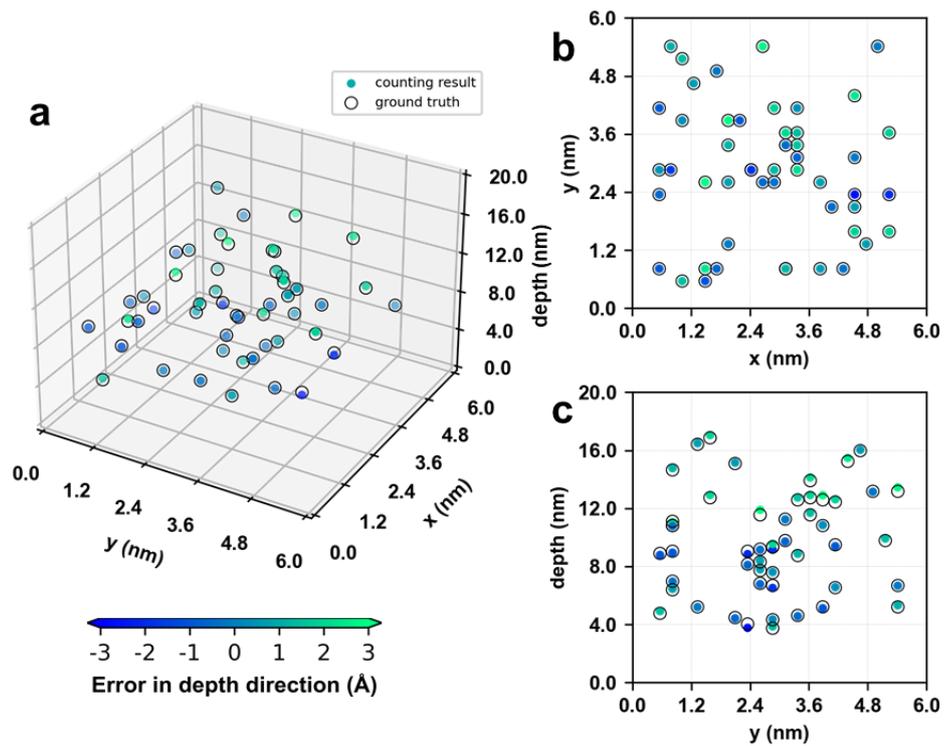

**Extended Data Fig. 3. Simulation test of positioning single interstitial atoms.** (a) 3D distribution of interstitial atoms determined in the APP phase images (color spots) and the true positions of the atoms in the model (black circles). The color of the spots represents the difference between the corresponding circle and spot, which is the positional error in the depth direction. (b) Projection of the atoms along the electron beam direction. (c) Projection of the atoms along the *x* direction.

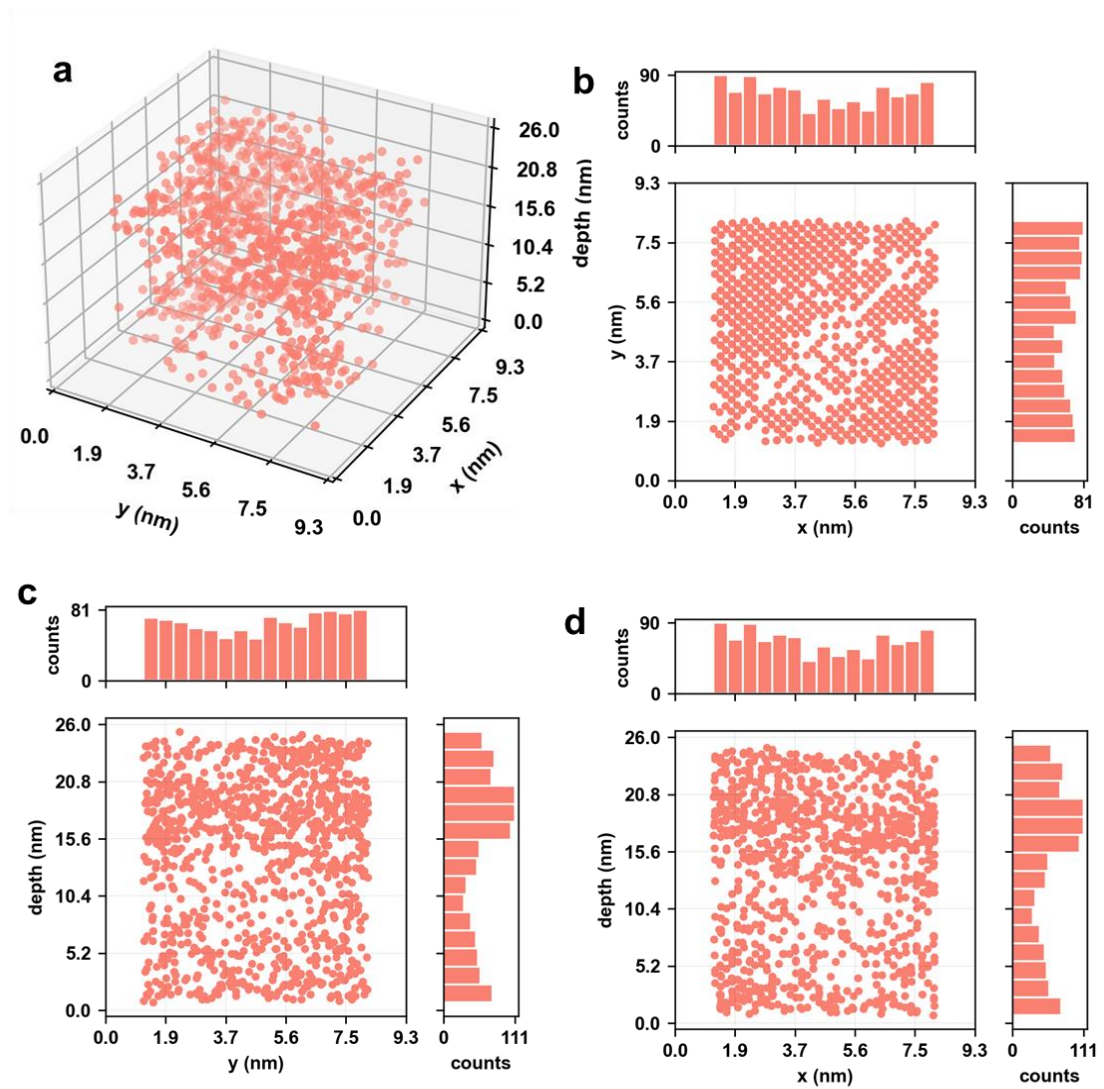

**Extended Data Fig. 4. Three-dimensional distribution of interstitial oxygen in the Mid-O region.** (a) The three-dimensional distribution of interstitial O atoms in Mid-O region. (b, c, and d) Projections along the three axes in (a). Each projection distribution graph contains a histogram of the number of interstitial atoms projected onto two coordinate axes.